\documentclass[10pt,letterpaper]{article}
\usepackage[top=0.85in,left=2.75in,footskip=0.75in]{geometry}

\usepackage{amsmath,amssymb}

\usepackage{changepage}

\usepackage{textcomp,marvosym}

\usepackage{cite}

\usepackage{nameref}

\usepackage[nopatch=eqnum]{microtype}
\DisableLigatures[f]{encoding = *, family = * }

\usepackage[table]{xcolor}

\usepackage{array}

\newcolumntype{+}{!{\vrule width 2pt}}

\newlength\savedwidth

\raggedright
\usepackage[aboveskip=1pt,labelfont=bf,labelsep=period,justification=raggedright,singlelinecheck=off]{caption}

\makeatletter
\renewcommand{\@biblabel}[1]{\quad#1.}
\makeatother

\usepackage{lastpage,fancyhdr,graphicx}
\usepackage{epstopdf}
\fancyheadoffset[L]{2.25in}
\fancyfootoffset[L]{2.25in}
\begin{document}
\vspace*{0.2in}

\begin{flushleft}
{\Large
\textbf\newline{Bounded distributions place limits on skewness and larger moments} 
}
\newline
\\
David J. Meer*,
Eric R. Weeks
\\
\bigskip
Department of Physics, Emory University, Atlanta, Georgia, USA
\bigskip

%
%





* dmeer@emory.edu (DM)

\end{flushleft}
\section*{Abstract}
Distributions of strictly positive numbers are common and can be characterized by standard statistical measures such as mean, standard deviation, and skewness.  We demonstrate that for these distributions the skewness $D_3$ is bounded from below by a function of the coefficient of variation (CoV) $\delta$ as $D_3 > \delta-1/\delta$.  The results are extended to any distribution that is bounded with minimum value $x_{\rm min}$ and/or bounded with maximum value $x_{\rm max}$.  We build on the results to provide bounds for kurtosis $D_4$, and conjecture analogous bounds exists for higher statistical moments.




\section{Introduction}
One often considers a probability function $P(x)$ of a random variable $X$. Distributions of $P(x)$ are characterized by quantities such as mean, median, standard deviation, and skewness. For a continuous random variable, $X$, $P(x)$ is the probability density of finding a value $x$ in the range $(x,x+dx)$.  For a discrete random variable, $P(x)$ is a discrete probability distribution which assigns a probability $p_i$ to each potential value $x_i$. The skewness is a measure of the asymmetry of a distribution \cite{bib1}.  While there are several possible definitions of skewness \cite{bib13}, a common definition depends on the third moment of the distribution compared to the second moment \cite{bib11,bib12}.  In particular, one can define the $n$th central moment for continuous or $N$ discreet variables as
\begin{eqnarray}
    \label{moments}
    m_n = \langle (x-\mu)^n \rangle & = & \int_{-\infty}^{+\infty} P(x) (x - \mu)^n  dx \\
    & = & \sum_{i=1}^{N} p_i(x_i - \mu)^n \nonumber
\end{eqnarray}
where $\mu$ is the mean of the distribution and $\sqrt{m_2}=\sigma$ is the standard deviation.  The standardized moments $D_n$ are defined as:
\begin{equation}\label{Dndef}
    D_n = \frac{m_n}{m_2^{n/2}}.
\end{equation}

We define skewness as the third standardized moment, $D_3$. This definition for skewness has the advantage that it is dimensionless.  It also has the useful property that distributions $P_1(x)$ and $P_2(x) = c P_1(cx + d)$ have the same skewness for $c>0$ and any $d$ \cite{bib13}.  Pearson \cite{pearson} derived an upper boundary on the skewness:
\begin{equation}
    D_3^2 \leq m_4 / m_2^2 - 1. \label{pearsonlimit}
\end{equation}
Alternate derivations of this result are also in the literature \cite{bib9,bib5}.
This applies for all distributions.

\begin{figure}
\includegraphics[bb= 22 141 621 369,width=0.95\textwidth]{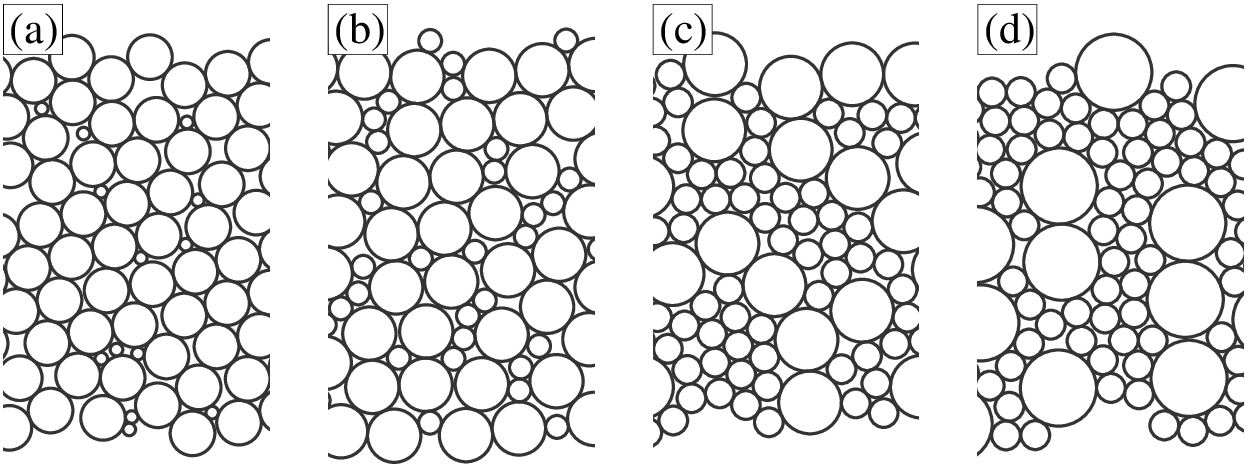}
\caption{\label{disks} Examples of circles with random bidisperse distributions of diameters with (a) CoV $\delta=0.4$, skewness $D_3=-1$; (b) $\delta=0.4, D_3=0$; (c) $\delta=0.4, D_3 = 1$; and (d) $\delta=0.4, D_3=+3$.  Decreasing the skewness requires the small circles becoming even smaller (compared to the mean size), as well as decreasing their frequency of occurrence.  The decreasing size reaches its natural limit when the small particles have zero size at $S=-2.1$ for $\delta=0.4$, as predicted in Eq (\ref{smolalski}).}
\end{figure}

Distributions of strictly positive numbers are often relevant:  numbers of objects, sizes of objects such as Fig \ref{disks}, ages of people, prices, barometer measurements, {\it etc}.  Such distributions have only non-negative support; one can more broadly consider distributions with bounded support, with boundaries $x_{\rm min}$ and/or $x_{\rm max}$, generically $x_{\rm bound}$.  Smo{\l}alski \cite{bib2} worked out upper and lower bounds on the skewness that applies for distributions with bounded support:
\begin{equation}
    \label{smolalski}
    {D_3}_{\rm min, max} = \delta_{\rm min,max} - \frac{1}{\delta_{\rm min,max}}
\end{equation}
with $\delta_{\rm min} = \sigma/(\mu - x_{\rm min})$ to determine ${D_3}_{\rm min}$ and $\delta_{\rm max} = \sigma/(x_{\rm max} - \mu)$
to determine ${D_3}_{\rm max}$.

In this paper, we present an alternative derivation for these skewness bounds.  Smo{\l}alski's derivation relies on the argument that achieving the extrema of skewness requires a bidisperse distribution.  We mathematically prove that this is indeed the case in Section \ref{skewsec}.  Smo{\l}alski then uses Lagrange multipliers to derive Eq (\ref{smolalski}); here, we use calculus to derive this equation and extend it to all real bounds. Our method also applies to higher order standardized moments, for which we find similar bounds in Section \ref{Dn}. We state the bounds, show when their behavior can be used to find the maximum or minimum standardized moment, ${D_n}_{\rm extr}$, and conjecture that these extrema apply to all distributions, not just bidisperse.

We are treating just the value of skewness corresponding to the parent distribution, rather than the sample skewness based on a finite number of samples which has different limits, see \cite{bib6}.  Note also that there are other definitions of skewness, for example that use the median of the distribution as part of the calculation \cite{bib1}, for which other limits exist \cite{bib10,bib4,bib8}.

\section{Results for Skewness} \label{skewsec}

We begin in the lowest order nontrivial case $n=3$, replicating Smo{\l}alski's skewness results. A distribution function with a low value of skewness has small values which rarely occur, for example the smallest circles seen in Fig \ref{disks}a.  A distribution with a high value of skewness is the opposite situation, where the large values rarely occur, for example the largest circles seen in Fig \ref{disks}d. For a distribution $P(x)$ with only non-negative support, the largest possible values of $x$ are unbounded, but the lowest possible values are bounded by zero.  Thus, it makes intuitive sense that the skewness will have a minimum possible value.

Our derivation will proceed by first considering bidisperse distributions with nonnegative support and showing that for a fixed $\delta$, the distribution with one value equal to zero achieves the lowest possible skewness.  We then show taking two distributions obeying Eq (\ref{smolalski}) and considering a weighted sum will result in a new distribution that also obeys Eq (\ref{smolalski}).  Next, we argue that any continuous distribution can be approximated by an appropriately weighted sum of bidisperse distributions.  In Sec. \ref{general} we will conclude by generalizing from distributions with non-negative support to distributions with arbitrary bounds, including those with $\mu\le 0$.

\subsection{Skewness for bidisperse distributions}

We start by considering a bidisperse distribution, $P(x)$ which takes on values $a_+,a_-$ with probabilities $q,p=1-q$.  Following \cite{desmond14}, we define the ratio
\begin{equation}
    \eta = a_+ / a_-
\end{equation}
and focus on $q$ as another important variable describing the distribution.  The meaning of the subscripts in $a_+$ and $a_-$ is the former is the value larger than the mean $\mu$ and the latter is smaller than $\mu$, respectively.  Knowing the mean $\mu$ allows us to relate these quantities as
\begin{eqnarray}\label{aandb}
    a_+ &=& \eta\mu / (1 - q + \eta q) \\
    a_- &=& \mu/(1 - q + \eta q).\nonumber
\end{eqnarray}
Note that a bidisperse distribution with a given $(\eta,q)$ is equivalent to a distribution with $(1/\eta,1-q)$ with swapped $a_+$ and $a_-$.  A key concept which we will use for much of this derivation is that in addition to the mean $\mu$, in general knowing any other two quantities related to the distribution will uniquely determine the distribution.  Those two quantities could be the values $a_+$ and $a_-$; they could be $\eta$ and $q$ as per Eq. \ref{aandb}.  Usefully, they can also be the standard deviation and skewness.  Thus, we will show that a distribution with $a_-$ achieving the minimum possible value ($a_- = 0$) is one where the skewness $D_3$ achieves its minimum value.

Given a bidisperse distribution defined as above, the standard deviation $\sqrt{m_2}=\sigma$ and skewness $D_3$ are then expressed as
\begin{eqnarray} \label{sigmagamma}
    \sqrt{m_2}=\sigma &=& \left((1-q)(a_--\mu)^2 + q (a_+-\mu)^2\right)^{1/2}\\
    D_3 &=& \frac{(1-q)(a_--\mu)^3 + q(a_+-\mu)^3}{m_2^{3/2}}.\nonumber
\end{eqnarray}

While $P(x)$ could be a distribution of a quantity with dimensions (such as a probability distribution of weights), our goal is to understand the non-dimensional skewness $D_3$.  Thus, rather than considering $\sigma$ which  has dimensions of $x$, we will use the non-dimensional quantity ``coefficient of variation,'' (CoV) defined as:
\begin{equation}
    \delta = \sigma / \mu.
    \label{cov}
\end{equation}
Here we use the symbol $\delta$ and later in this manuscript we will generalize this symbol beyond the specific meaning of CoV. We can use Eqs (\ref{aandb}) to eliminate $a_+$ and $a_-$ from $m_2$ and $D_n$, resulting in
\begin{eqnarray}
    \frac{\sqrt{m_2}}{\mu} = \delta &=& \frac{(\eta-1)\sqrt{q-q^2}}{1-q + \eta q}\label{sigmaeqn}\\
    D_3 &=& \frac{2 q - 1}{\sqrt{q - q^2}}.\label{gammaeqn}
\end{eqnarray}
 These require $\eta>1$. Eqs. (\ref{sigmaeqn},\ref{gammaeqn}) can be inverted to provide expressions for $q$ and $\eta$ in terms of $\delta$ and $D_3$. We include the substitution $M_3=\sqrt{4+D_3^2}$ which will be a reoccurring term:
\begin{eqnarray}
    q &=& \frac{\pm D_3 + M_3}{2M_3}\label{qfuncD3}\\
    \eta &=& \frac{2-\delta(\pm D_3 -M_3)}{2-\delta(\pm D_3 + M_3)}.\label{etafuncD3}
\end{eqnarray}

These two equations give rise to two branches of solutions depending on whether the $+$ or $-$ is taken in each equation.  Inspection shows that the negative sign in Eqs (\ref{qfuncD3},\ref{etafuncD3}) arrives back at the classical definition of skewness, whereas the positive branch has no significance.  For the remainder of our consideration of $D_3$, we will use the negative branch of the solutions and drop the $\pm$ symbol. We continue and calculate the two possible values according to Eq (\ref{aandb}):
\begin{eqnarray}
    a_+ &=& \mu\left(1+\frac{\delta}{2}\left(D_3+M_3\right)\right)\label{bD3}\\
    a_- &=& \mu\left(1+\frac{\delta}{2}\left(D_3-M_3\right)\right).\label{aD3}
\end{eqnarray}

Using Eq (\ref{aD3}), we can do a straightforward calculation for the minimum possible skewness $D_3(\delta)$ for bidisperse distributions with $a_+,a_- \geq 0.$  A distribution with a low skewness is one that has a small amount of small numbers:  and the smallest number we can get for a distribution of strictly non-negative numbers is zero.  Thus, to find the limit on skewness, we solve Eq (\ref{aD3}) for $a_-=0$. This also implies $a_+=\mu/(1-q)$. Solving for $D_3$ when $a_-=0$ in Eq (\ref{aD3}) lets us solve for ${D_3}_{\rm min}$:
\begin{equation}\label{smolalski2}
    {D_3}_{\rm min} = \delta - \frac{1}{\delta}.
\end{equation}
For example, this gives values ${D_3}_{\rm min} = -2.1$ for $\delta=0.4$, and ${D_3}_{\rm min}=0$ for $\delta = 1$.

\subsection{A bidisperse distribution with $a_{\rm min} > 0$ increases $D_3$}
\label{Deltaasection}

For a fixed value of $\delta$, if the minimum value of the distribution $a_-$ is larger than zero, then $D_3$ will increase.  This is not straightforward to see from the equations above, but an alternate formulation will work.  Define:
\begin{eqnarray}
    \Delta_+' &=& a_+-\mu > 0\\
    \Delta_-' &=& \mu-a_- > 0\nonumber
\end{eqnarray}
Using Eqs. (\ref{aandb}) we can factor out $\mu$ and arrive at normalized definitions of $\Delta_{+,-}=\Delta_{+,-}'/\mu$ We then have the probability of $a_+$ being
\begin{equation}
    q = \frac{\Delta_-}{\Delta_- + \Delta_+}.
\end{equation}
We can then get $\delta$ using
\begin{equation}
     \delta^2 = (1-q)\Delta_-^2 + q \Delta_+^2 = \Delta_- \Delta_+.
\end{equation}
Given that we wish to keep $\delta$ constant, we can thus use $\Delta_+ = \delta^2/\Delta_-$ to eliminate $\Delta_+$, leading to
\begin{equation}
    q = \frac{1}{1+\Delta_-^2/\delta^2}.
\end{equation}
Now consider the third moment of the distribution $m_3$:
\begin{eqnarray}
            m_3 &=& \mu^3\left((1-q) \Delta_-^3 - q \Delta_+^3\right)\\ \nonumber
         &=& \mu^3\delta^2 (\Delta_- - \Delta_+) \\ \nonumber
         &=& \mu^3\delta^2 (\frac{\delta^2}{\Delta_-} - \Delta_-).\nonumber
\end{eqnarray}

The partial derivative of $m_3$ with respect to $\Delta_-$ holding $\delta$ constant is
\begin{equation}
        \left( \frac{\partial m_3}{\partial \Delta_-} \right)_\delta  = -\mu^3\left(\frac{\delta^4}{\Delta_-^2} +\delta^2\right) < 0.
\end{equation}
Increasing $\Delta_-$ always decreases $m_3$, assuming we keep $\delta$ constant and $\mu$ positive.  Likewise, decreasing $\Delta_-$ (making $a_-$ larger than zero) will always increase $m_3$.  Thus, making $a_-$ larger than zero must increase the skewness $D_3$.  This proves that for the bidisperse distribution with a fixed $\delta$, Eq (\ref{smolalski2}) is indeed the lowest possible skewness.

\subsection{Generalizations of skewness results}

Suppose we have two separate distributions $P_r(x)$ and $P_s(x)$ both with mean $\mu$ and both satisfying the bound of Eq (\ref{smolalski2}).  We wish to show that any combination of these two distributions, $P_t(x) = \alpha P_r(x) + (1-\alpha) P_s(x)$ (with $0 \leq \alpha \leq 1$), also satisfies Eq (\ref{smolalski2}).  Given that the means are identical, it is straightforward that $\delta^2_t = \alpha \delta^2_r + (1-\alpha) \delta^2_s$ and also $m_{3,t} = \alpha m_{3,r} + (1-\alpha) m_{3,s}$.  As $D_3 = m_3 / m_2^{3/2}$, we can rewrite the bound on skewness Eq (\ref{smolalski2}) as
\begin{equation}
    m_{3,{\rm min}} \geq \mu^3\left(\delta^4 - \delta^{2}\right)
    \label{combinbound1}
\end{equation}
Given that both $P_r$ and $P_s$ satisfy this constraint, we have
\begin{eqnarray}
     m_{3,r} &\geq& \mu^3(\delta_r^4 - \delta_r^2)\\
     m_{3,s} &\geq& \mu^3(\delta_s^4 - \delta_s^2),\nonumber
\end{eqnarray}
and thus
\begin{eqnarray}
    m_{3,t} &=& \alpha m_{3,r} + (1-\alpha) m_{3,s}\label{result1}\\
    &\geq& \mu^3\left(\alpha (\delta_r^4 - \delta_r^2) + (1-\alpha)(\delta_s^4 - \delta_s^2)\right)\nonumber\\
    &=& \mu^3\left(\alpha \delta_r^4 + (1-\alpha) \delta_s^4 - \delta_t^2\right), \nonumber
\end{eqnarray}
where the last line uses the expression for $\delta^2_t$ introduced above.  Next, note that
\begin{eqnarray}
    \delta_t^4 &=& (\alpha \delta_r^2 + (1-\alpha) \delta_s^2)^2\label{result2}\\
    &=& \alpha^2 \delta_r^4 + 2 \alpha (1-\alpha) \delta_r^2 \delta_s^2 +
        (1-\alpha)^2 \delta_s^4.\nonumber
\end{eqnarray}
On the right-hand side of Eq (\ref{result1}), add $\mu^3\delta_t^4$ and subtract the right-hand side of Eq (\ref{result2}):
\begin{eqnarray}
m_{3,t} &\geq&\mu^3(\alpha \delta_r^4 + (1-\alpha) \delta_s^4 - \delta_t^2+\delta_t^4\\
    & &-\alpha^2 \delta_r^4 - 2 \alpha (1-\alpha) \delta_r^2 \delta_s^2 -
        (1-\alpha)^2 \delta_s^4) \nonumber
\end{eqnarray}
Every term without $\delta_t$ on the right-hand side can be combined as $\mu^3\alpha(1-\alpha)(\delta_r^2 - \delta_s^2)^2$ which is always non-negative, so thus
\begin{equation}
    m_{3,t} \geq \mu^3(\delta_t^4 - \delta_t^2),
    \label{combinbound2}
\end{equation}
proving that the combined distribution function $P_t(x)$ must satisfy Eq (\ref{smolalski}) if the two original distributions satisfy that bound.

Finally, we need to generalize from the bidisperse distribution to any distribution.  Following \cite{bib9}, we observe that any continuous distribution with some fixed $\mu=\mu_0$ can be approximated by a discrete distribution with values $a_i$ and probabilities $p_i$ and $\mu=\mu_0$.  Rohatgi and Sz{\'e}kely then proved that any such discrete distribution can be decomposed into a sum of discrete distributions with two values and $\mu=\mu_0$, that is, the bidisperse distributions that we have been considering (see also  Appendix A).  In the previous paragraph, we have shown that sums of distributions satisfy the bound.  Thus, we have proven that Eq (\ref{smolalski}) holds for any distribution $P(x)$ of strictly non-negative values of $x$.

\subsection{Distributions bounded by $x_{\rm min}$ or $x_{\rm max}$}
\label{general}

We have considered distributions $P(x)$ for which $x \geq 0$. By rescaling the distribution, we can enforce any value of $\mu$ we would like. However, this comes at the expense of potentially running into our bounds. For example, you cannot have some $\mu\le 0$ without a minimum less than or equal to zero. When some values of $x$ are below 0, we cannot simply rescale by a constant multiple to enforce the bounds.  Of course, an additive constant would fix a distribution and make it non-negative.  As noted in the introduction, this also leaves $D_3$ unchanged:  consider $P(x)$ and $P'(x) = P(x - d)$.  $\mu' = \mu+d$ but as the moments are defined as $\langle (x-\mu)^n \rangle$, $m_2$ and $m_3$ are unchanged by this shift, and thus $D_3$ does not change.

Similarly, we also note that $\lim_{\mu\to 0^+}(a_+,a_-,\eta)=\lim_{\mu\to 0^-}(a_+,a_-,\eta)$. This limit can be calculated directly by multiplying by $\frac{\mu}{\mu}$ in Eq (\ref{etafuncD3}) and distributing the $\mu$ factor in Eqs (\ref{bD3},\ref{aD3}), leaving us with just $\sqrt{m_2}$ where there was previously $\delta$. Therefore, we do not have to be concerned with means approaching zero.

Now consider the general case of a distribution $P(x)$ bounded by $x_{\rm min}$ from below and with a mean $\mu$ which might be zero.  Let us assume $P(x)$ has a nontrivial domain, which is to say, it is not a distribution which is only nonzero at one value (which would thus be $\sigma=0, D_3=0$).  The transformed distribution $P'(x) = P(x + x_{\rm min})$ has mean $\mu'=\mu-x_{\rm min}$.  This transformed distribution now is nonzero only for $x \geq 0$, so is one of the distributions we considered above, and since the distribution has a nontrivial domain, $\mu' > 0$ must be true.  Therefore, we have:
\begin{equation}
    \label{deltaxmin}
    \delta = \sigma/(\mu - x_{\rm min}).
\end{equation}
That is, $\delta$ depends on the standard deviation $\sigma$ and mean $\mu$ of the original distribution $P(x)$, with the additional correction of subtracting $x_{\rm min}$, at which point we can use Eq (\ref{smolalski2}) to find ${D_3}_{\rm min}$.

The other interesting case is a distribution bounded by $x_{\rm max}$ from above.  Considering $P''(x) = P(-x)$ changes the mean to be $\mu'' = -\mu$ and the skewness to be $D_3'' = -D_3$, but does not change the standard deviation.  The distribution $P''(x)$ is now bounded from below by $-x_{\rm max}$ so we get:
\begin{equation}
    \label{deltaxmax}
    \delta = \sigma / (x_{\rm max} - \mu),
\end{equation}
which goes into Eq (\ref{smolalski2}) to calculate ${D_3}_{\rm min}$.  In this case, we actually have found ${D_3}_{\rm max} = -{D_3}_{\rm min}$.  Thus, we have rederived the results of \cite{bib2}, that is, Eq (\ref{smolalski}).

If a distribution $P(x)$ has domain $x_{\rm min} \leq x \leq x_{\rm max}$ then the above results give both a lower and an upper bound on $D_3$.  As a conceptual example, suppose that $x_{\rm min} = \mu-3 \sigma$ and $x_{\rm max} = \mu+3 \sigma$; then $-8/3 \leq D_3 \leq 8/3$.  This is consistent with the empirical observation that the skewness tends to lie between -3 and +3.

As a useful check on these results, consider the bidisperse distribution again with probability $P(a_+)$ and $P(a_-)$ for sizes $a_- < a_+$.  Here we have $x_{\rm min}=a_-$, and CoV given by Eq (\ref{sigmaeqn}).  Using Eqs (\ref{aandb}), (\ref{deltaxmin}), and (\ref{smolalski}), one can solve for ${D_3}_{\rm min}$ in terms of the variables $\eta$ and $q$, recovering Eq (\ref{gammaeqn}):  that is, ${D_3}_{\rm min}$ is achieved in this situation.  Similarly, using $x_{\rm max}=a_+$ one finds again ${D_3}_{\rm max}=D_3$.

If we extend Eq (\ref{smolalski2}) to any arbitrary upper or lower bound $x_{\rm bound}$, we get the following relationship for the extreme value of $D_3$, ${D_3}_{\rm extr}$
\begin{equation}
    {D_3}_{\rm extr} = \frac{\delta}{1-x_{\rm bound}/\mu} - \frac{1-x_{\rm bound}/\mu}{\delta}\label{D3x}
\end{equation}
which has reprised Eq (\ref{smolalski}).

\section{Extensions to higher order moments} \label{Dn}

\subsection{Notes to Generalize from Skewness \label{ageneric}}

Going forward, we note that Eq (\ref{D3x}) is useful for more than the extreme $D_3$ of the system, when considering a bidisperse system.  As noted at the start of Sec. \ref{skewsec}, if one is given $\mu$ and two other quantities, then one can uniquely determine a bidisperse distribution.  Thus knowing one size $x_{\rm bound}$, $\delta$, and $\mu$, determines the other size and relative probabilities.  By plugging in any generic size $a/\mu$, which could be $a_+/\mu$ or $a_-/\mu$ to Eq (\ref{D3x}), this produces the $D_3$ that makes a bidisperse distribution with that size and a given CoV $\delta$. This equation can be solved for $a$ to give either of Eqs. (\ref{bD3},\ref{aD3}). In other words, if we know we have a bidisperse distribution, then Eq (\ref{D3x}) is a formula for $D_3$ as a function of one of the sizes $a$.  We will derive similar results for higher moments.

\subsection{Kurtosis $D_4$}

As noted in the introduction, previous results by Pearson \cite{pearson} show that $D_4\ge D_3^2+1$ for any given distribution as per Eq (\ref{pearsonlimit}).  If we now know an inequality for $D_3$ on any distribution with Eq (\ref{D3x}), we can solve for a new limit in $D_4$. in terms of $x_{\rm min}$, $\mu$, and $\delta$.  In particular, we have to consider two cases.  Treating the situation where the distribution has only nonnegative support ($x_{\rm min} = 0)$, then for $\delta < 1$, $D_{3,\rm min} < 0$.  This implies that $D_3 = 0$ is also possible, and therefore we can achieve lower $D_4$ than is predicted by Eq (\ref{pearsonlimit}) based on $D_{3, \rm min}$.  In other words, we can consider the bidisperse distribution with $D_3=0$, which can be found using Eqs (\ref{qfuncD3},\ref{etafuncD3}), to achieve $D_{4,\rm min} = 1$ as per Eq (\ref{pearsonlimit}).  For $\delta \geq 1$, $D_{3,\rm min} \geq 0$ and the limit on $D_4$ then follows from Eq (\ref{D3x}).  Thus we have
\begin{align}
    D_4 & \geq 1 &(\delta < 1) \nonumber \\
    D_4 & \geq \left( \delta - \frac{1}{\delta} \right)^2 + 1 &(\delta \geq 1) \label{D4pearson}
\end{align}
for the limits on $D_4$ in the two cases.

For the more general case of a distribution bounded on one side (by either $x_{\rm min}$ from below, or $x_{\rm max}$ from above, but not both), we can define the limits on kurtosis $D_4$ in terms of the extremum bounding value $x_{\rm extr}$.  Define
\begin{equation}
    \delta_0 = \left|1 - x_{\rm extr}/\mu \right| .
    \label{delta0}
\end{equation}
That is, $\delta_0$ is the equivalent of Eqs (\ref{deltaxmin},\ref{deltaxmax}).
We then get
\begin{align}
    D_4 & \geq 1 {} &(\delta < \delta_0) \\
    D_4 & \geq \left(  \frac{\delta}{\delta_0} -\frac{\delta_0}{\delta} \right)^2 + 1 &(\delta \geq \delta_0). \label{D4general}
\end{align}
In other words, whether the distribution is bounded from below or bounded from above, in both cases this sets a minimum on $D_4$ -- but not a maximum.

When the distribution is bounded from below by $x_{\rm min}$ and bounded from above by $x_{\rm max}$, the situation complicates further.  We start by defining $\delta_{\rm min}$ and $\delta_{\rm max}$ analogously to Eq (\ref{delta0}).  While $x_{\rm min} < x_{\rm max}$, the ordering of $\delta_{\rm min}$ and $\delta_{\rm max}$ is not determined.  Thus define
\begin{align}
    \delta_1 & = {} \min(\delta_{\rm min}, \delta_{\rm max}),\\
    \delta_2 & = {} \max(\delta_{\rm min}, \delta_{\rm max}),\\
    D_{4,m}(\delta) & = {} \left(  \frac{\delta}{\delta_m} -\frac{\delta_m}{\delta} \right)^2 + 1,
\end{align}
where $m=1,2$.  Next define $\delta'$ using
\begin{equation}
    D_{4,1}(\delta') = D_{4,2}(\delta')
\end{equation}
which can be solved to get $\delta' = \sqrt{\delta_1 \delta_2} =\sqrt{\delta_{\rm min} \delta_{\rm max}}$.  The limits on kurtosis $D_4$ are then
\begin{align}
    1 \leq & D_4 \leq D_{4,2}(\delta) &(\delta < \delta_1) \nonumber \\
    D_{4,1}(\delta) \leq & D_4 \leq D_{4,2}(\delta) &(\delta_1 \leq \delta < \delta') \label{D4mostgeneral}
\end{align}
and values $\delta > \delta'$ are disallowed as they would require the bidisperse distribution be composed of values that lie outside of one or both of the boundaries $(x_{\rm min},x_{\rm max})$.  At $\delta = \delta'$, the only bidisperse distribution that is valid is composed of the two values $(x_{\rm min}, x_{\rm max})$ with appropriate probabilities necessary to get the value of $\delta$, and we have $D_{4,1} = D_{4,2} = D_4$.

\begin{figure}
    \centering
    \includegraphics[width=0.49\textwidth]{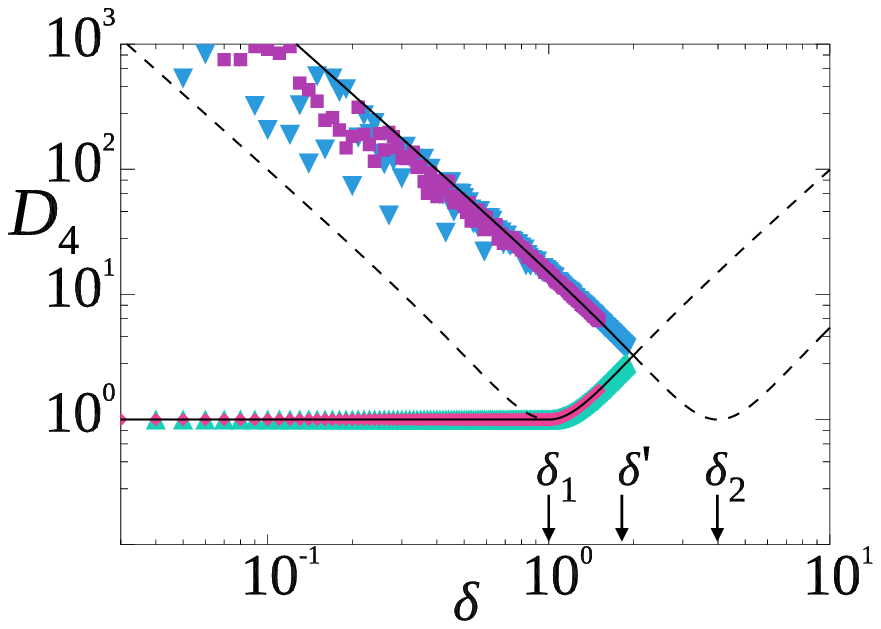}
    \includegraphics[width=0.49\textwidth]{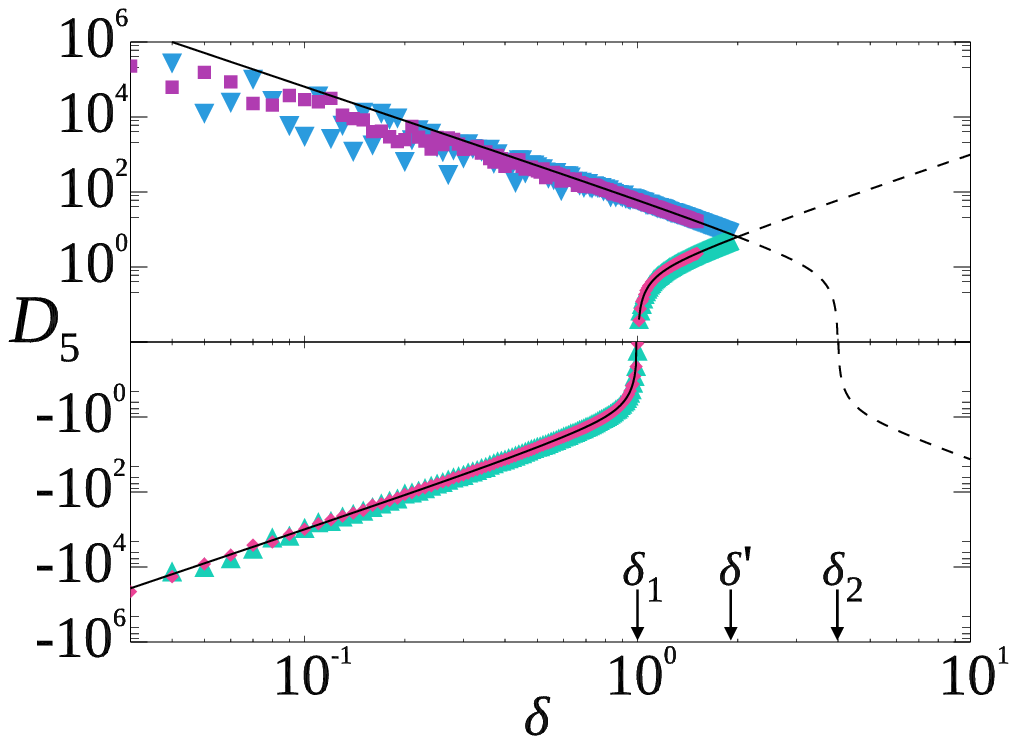}
    \caption{Simulations with bidisperse distributions, and tri- or quad-disperse distributions, yield extrema which are plotted against the prediction (black line) given by Eq. (\ref{DNlimits}) for $D_4$ (left) and $D_5$ (right).  The data correspond to distributions with values limited to be greater than zero and less than 5$\mu$. The bidisperse triangles are green (pointing up) for the minima and blue (pointing down) for the maxima, and the tri or quad-disperse are pink (diamonds) for minima and purple (squares) for maxima. The more extreme values from quad or tri-disperse was plotted for each polydispersity bin.}
    \label{fig:numerical}
\end{figure}

These results are visualized in Fig \ref{fig:numerical}a, which illustrates a specific example with $x_{\rm min}=0, x_{\rm max} = 5,$ and $\mu=1$.  For this example, $\delta_1=1.0$ and $\delta'=3.25$.  The solid lines indicate Inequalities \ref{D4mostgeneral}, and the symbols indicate simulated random distributions with a specified $\delta$.  Specifically, we generated distributions with data lying between limits $x_{\rm min}, x_{\rm max}$, and with enforced mean $\mu$, and calculated $\delta$ and $D_4$ for all.  For a given small range of $\delta$, we generated 20,000 distinct random distributions, half that are bidisperse, and the other half with three or four values.  Over these 20,000 distributions, Fig \ref{fig:numerical}a plots the maximum and minimum $D_4$ found for each $\delta$, all of which lie between the limits corresponding we have found (shown by the lines).  While we have not proven that the bidisperse distribution sets the limits for $D_4$ for all other distributions, this is suggestive that Inequalities \ref{D4mostgeneral} are indeed limits for the kurtosis for any distribution.

\subsection{Higher order generalized moments}
\label{higherorder}

We now proceed with an alternate derivation of Inequalities \ref{D4general} which we can extend to higher moments.  The generic definition of $D_n$ in the bidisperse case is:
\begin{equation}
D_n=\frac{(1-q)(a_{-} - \mu)^n+q(a_{+} - \mu)^n}{m_2^{n/2}}.\label{Dnbi}    
\end{equation}
If we use Eqs (\ref{aandb}) to solve for the generic definition of $D_n$ in terms of $q$ and $\eta$, we arrive at a formula of only $q$:
\begin{equation}
    D_n = \frac{\left(1-q\right)^{n-1}+\left(-1\right)^n q^{n-1}}{\left(q-q^{2}\right)^{\frac{n}{2}-1}}. \label{Dnrhoeqn}
\end{equation}
Plugging in $n=3$ arrives back at Eq (\ref{gammaeqn}). 

For a bidisperse distribution, we can rewrite the second line of Inequality \ref{D4general} as an equality in terms of $a$, one of the two bidisperse values.  We then note that Eqs (\ref{D3x},\ref{D4general}) are both functions of $z = \delta / (1 - a/\mu)$:
\begin{eqnarray}
    D_3 &=& \frac{\delta}{1-a/\mu} - \frac{1-a/\mu}{\delta} = z^1-z^{-1}\\
    D_4 &=& \frac{\delta^4-\delta^2(1-a/\mu)^2+(1-a/\mu)^4}{\delta^2(1-a/\mu)^2}\nonumber\\
    &=& z^2 - z^0 + z^{-2}. \nonumber
\end{eqnarray}
(In $D_4$, because only even powers of $z$ appear, the absolute value signs in Eq (\ref{delta0}) can be dropped, allowing $z$ to have the same meaning for both $D_3$ and $D_4$.)
The general pattern appears to be a finite sum of a geometric series.  In fact, Appendix \ref{Dnappendix} shows how one can start from Eq (\ref{Dnrhoeqn}) to derive 
\begin{eqnarray}
    D_n &=& \sum_{i=1}^{n-1}\left(-1\right)^{(n-i+1)}\left(\frac{\delta}{1-a/\mu}\right)^{2i-n}
    \label{DNlimits} \\
    &=& \sum_{i=1}^{n-1}\left(-1\right)^{(n-i+1)} z^{2i-n} . \nonumber
\end{eqnarray}
One can immediately put in a value for $a$ of interest and get a potential limit of $D_n$.  For example, for distributions bounded from below by $x_{\rm min}$ we conjecture
\begin{equation}
    D_5 \geq D_{5,{\rm min}} = z^3 - z^1 + z^{-1} - z^{-3}
\end{equation}
with $z = \delta / (1 - x_{\rm min}/\mu)$ as above.  As with $D_3$, our conjectured $D_{5,{\rm max}}$ is a similar equation using $z = \delta / (x_{\rm max}/\mu - 1)$.  Figure \ref{fig:numerical}b shows these two limits as the solid lines for the case $x_{\rm min}=0$, $x_{\rm max}=5$, and $\mu=1$, along with the maximum and minimum observed $D_5$ values from numerically generated random distributions.  All the random distributions lie within our conjectured analytic limits, again suggestive that they are the actual limits.

To try to show that these bounds achieve minima for any $n$, we can try a similar method as section \ref{Deltaasection}. If we write out a more generic $m_n$:
\begin{equation}
m_n=\frac{\Delta_-^2}{\Delta_-^2+\delta^2}\left(\frac{\delta^2}{\Delta_-}\right)^n+\left(1-\frac{\Delta_-^2}{\Delta_-^2+\delta^2}\right)(-\Delta_-)^n
\label{Deltmn}\end{equation}
we then can take its derivative with respect to $\Delta_-$, giving
\begin{eqnarray}
\frac{\partial m_n}{\partial \Delta_-}=\frac{\left(-1\right)^{n}\left(\delta^{2}\Delta_{-}^{n-1}\right)\left(\left(n-2\right)\Delta_{-}^{2}+n\delta^{2}\right)}{\left(\Delta_{-}^{2}+\delta^{2}\right)^{2}} \label{derivmn}\\
-\frac{\delta^{2n}\Delta_{-}^{1-n}\left(\left(n-2\right)\delta^{2}+n\Delta_{-}^{2}\right)}{\left(\Delta_{-}^{2}+\delta^{2}\right)^{2}}. \nonumber
\end{eqnarray}

Eq (\ref{derivmn}) is negative for all odd values of $n$, implying an increase in the smallest size above zero will only increase $D_n$:  thus, for odd $n$, $D_n$ is minimized for a bidisperse distribution with the smallest size set to zero.  For even $n$, negative values of Eq (\ref{derivmn}) are achieved for $\Delta_{-}$ between 0 and $\delta$, but positive for $\Delta_{-} > \delta$.  Thus, the minimum $m_n$ is achieved at $\Delta_{-} = \delta$.  In fact, this recapitulates the result of Eq (\ref{D4pearson}), that $D_{4,{\rm min}}$ is not a universal formula but rather depends on $\delta$.  Furthermore, if we try to replicate Eqs. (\ref{combinbound1}-\ref{combinbound2}) with $m_4$, the statements are untrue even when $\delta_r=\delta_s$. This gives credence that the boundaries of $D_n$ for even $n$ are not always given by the choice of $x_{\rm bound}$.

Lastly, as previously noted, a bidisperse distribution can be completely described by three parameters:  most directly by the values $a_-$, $a_+$, and the probability $q$ for one of these values.  Our approach has been to instead use $\mu$, $\delta$, and $a_-$ to find a constraint on $D_n$.  We note that Eq (\ref{DNlimits}) and the definition of $z$ is sufficient to find analogues of Eqs (\ref{qfuncD3}-\ref{aD3}):  thus, to use $D_n$, $\mu$, and $\delta$ to describe a bidisperse distribution.  One can start with those three quantities and determine $a_-$, $a_+$, and $q$: analytically for $D_3$ as per Eqs (\ref{qfuncD3} - \ref{aD3}), and numerically in other cases.  This has been useful in the past for finding distributions with desired values of the moments \cite{desmond14}.  Moreover, by then considering which values of $a_-$ and $a_+$ lie within bounds, one has a slightly alternate approach to finding bounds on $D_n$.

\section{Conclusion}
We have presented an alternative derivation of Eq (\ref{smolalski}) to that presented in \cite{bib2}; this equation provides bounds on the skewness $D_3$ for a bounded distribution with a given CoV $\delta$.  Equivalently, if $D_3$ is given, then this equation provides a bound for $\delta$.  Returning to our starting example, if one is considering a distribution of strictly positive numbers, then for a given $D_3$, Eq (\ref{smolalski}) can be solved for the maximum possible $\delta$.

Our results for $D_3$ naturally imply limits on $D_4$ (Inequalities \ref{D4general} using Pearson's formula \cite{pearson}, and Inequalities \ref{D4mostgeneral} more generally).  Our general methodology is to note that bidisperse distributions are characterized by three parameters, which most naturally are the two values $a_+$ and $a_-$ as well as the probability $q$ of the value $a_+$; however, one can fruitfully choose as the three parameters the mean $\mu$, coefficient of variation $\delta$, and $a_-$.  Setting $a_-$ to the lower bound of all possible distributions with a given $\mu$ and $\delta$ leads to lower bounds for $D_3$ and $D_4$.  Moreover, our methodology extends to higher moments, leading to conjectures for limits on higher standardized moments as discussed in Section \ref{higherorder}.  One possible extension to our work would be to see if there are other relationships between general $D_n$ and $D_m$.  It would also be interesting to discover a counterexample where a distribution exists that exceeds the limits of $D_n$ set by considering bidisperse distributions as in Section \ref{higherorder}.  We note that numerically at least, we have not found such a counterexample for $n=5$, as seen in the data of Fig \ref{fig:numerical}.

Our results have implications for a prior computational study of the packing of spheres, and how the density of such packings depend on the CoV and skewness of a particle size distribution \cite{desmond14}.  In that prior work, the results had a varying range of skewnesses but the authors did not comment on the choice of this range.  In fact, the lower bound on skewness studied in that work corresponds to result of Eq (\ref{smolalski}).  This bound implies that a sphere packing composed of a distribution of radii with a given $\delta$ and lowest possible skewness is, in fact, equivalent to the packing of a distribution of equal-sized spheres; and the observed density of such packings in \cite{desmond14} obeyed this property, as it must.  This is somewhat analogous to the circle packing shown in Fig.~\ref{disks}a, for which the skewness has not yet reached the lower limit; nonetheless the packing is dominated by circles of the larger size.  


%
%
%
\begin{appendix}
\section{Discreet Distribution Decomposition} 

Rohatgi and Sz{\'e}kely derived the result that any discrete distribution with mean $\mu$ can be decomposed into a sum of bidisperse distributions, all with mean $\mu$ \cite{bib9}.  Their derivation is terse, so we rederive the result in this Appendix with a slightly lengthier presentation.

First, consider a discrete distribution $P(x)$ where $x$ can take values $a_i$ with probability $p_i$ for $1 \leq i \leq n$, $\Sigma_i p_i=1$, and with mean $\Sigma_i p_i a_i = \mu$.  Replace $a_n$ and $a_{n-1}$ by
\begin{equation}
    \label{rohatgi}
    a'_{n-1} = \frac{p_{n-1}}{p_{n-1}+p_n} a_{n-1} + \frac{p_n}{p_{n-1}+p_n} a_n
\end{equation}
which occurs with probability $p'_{n-1} = p_{n-1}+p_{n}$.  This is now a new distribution with mean $\mu$ and one fewer value.  This can be repeated until one ends with a final distribution that takes on three discrete values, $a_1, a_2,$ and $a'_3$ with probabilities $p_1, p_2,$ and $p'_3$.

If we have a tridisperse distribution with three discrete values $(a_1, a_2, a_3),$ with probabilities $(p_1,p_2,p_3)$ and mean $\mu$, we can decompose this into the sum of two bidisperse distributions as follows.  Without loss of generality, assume $a_1 < \mu$ and $a_2 \leq \mu$.  Then the first bidisperse distribution has values $(a_1,a_3)$ with probabilities $p'_1=\frac{a_3-\mu}{a_3-a_1}$ and $p'_3=\frac{\mu-a_1}{a_3-a_1}$, and similarly for the second distribution with values $(a_2,a_3)$.  Sampling the first distribution with probability $p_1/p'_1$ and the second with probability $p_2/p'_2$ recovers the original tridisperse distribution.

Now consider the distribution with four discrete values $(a_1,a_2,a_3,a_4)$ and the related distribution $(a_1,a_2,a'_3)$ formed using Eq (\ref{rohatgi}).  The latter can be decomposed as a sum of two bidisperse distributions, as just demonstrated.  This then provides a scheme to reduce the four-valued distribution to a sum of two three-valued distributions, one of which eliminates $a_1$ and the other which eliminates $a_2$.  That is, the probability of finding $a'_3$ in each of the two bidisperse distributions is used to determine the new probabilities of finding $a_3$ and $a_4$ in the two tridisperse distributions.  Proceeding by induction, each distribution with $n$ distinct $a_i$ values can be decomposed into two distributions of $n-1$ distinct values, ultimately reducing down to a sum of bidisperse distributions.

\section{Derivation of Eq (\ref{DNlimits}) \label{Dnappendix}}

We wish to show that Eqs (\ref{Dnrhoeqn}) and (\ref{DNlimits}) are equivalent expressions for $D_n$ for a bidisperse distribution.
It is easiest to start with the end result and work backwards: Eq (\ref{DNlimits}) is
\begin{align}
    D_n & = {} \sum_{i=1}^{n-1}\left(-1\right)^{(n-i+1)}\left( z \right)^{(2i-n)}\label{DNlimitsrepriseZ}\\
    & = {} \sum_{i=1}^{n-1}\left(-1\right)^{(n-i+1)}\left(\frac{1-a/\mu}{\delta}\right)^{(n-2i)}\label{DNlimitsreprise}
\end{align}
where $a$ can represent either $a_+$ or $a_-$.  We will begin by examining the term with $a,\mu$, and $\delta$ and work to express it in terms of $\eta$ and $q$.  We will initially assume $a=a_-$ and use Eq (\ref{aandb}) to express $a_-$ in terms of $\eta, \mu$, and $q$; and likewise we will use Eq (\ref{sigmaeqn}) to express $\delta$ in terms of those same variables.  This leads to
\begin{align}
    \frac{1-a_{-}/\mu}{\delta} & = {} \frac{\mu-a_{-}}{\mu\delta}\\
    & = {} \frac{\mu - [\mu/(1-q+\eta q)]}{\mu (\eta-1) \sqrt{q-q^2}/(1-q+\eta q)}\\
    & = {} \frac{1-q+\eta q - 1}{(\eta-1)\sqrt{q-q^2}}\\
    & = {} \frac{q}{\sqrt{q(1-q)}} = \left( \frac{q}{1-q} \right)^{1/2}. \label{aminus}
\end{align}
We can put this in to Eq (\ref{DNlimitsreprise}) to give
\begin{align}
    \label{DNpart2}
    D_n & = {} \sum_{i=1}^{n-1}\left(-1\right)^{(n-i+1)}\left(\frac{q}{1-q}\right)^{(n/2)-i}\\
    & = {} \left(-1\right)^{n+1} \left(\frac{q}{1-q}\right)^{n/2}  \left[ \sum_{i=1}^{n-1}\left(-1\right)^i \left(\frac{1-q}{q}\right)^i \right] 
    \label{appendixa2}
\end{align}
where now the summation is simply a finite geometric sum.  The sum can be evaluated as
\begin{align}
    \sum_{i=1}^{n-1} \left(\frac{q-1}{q}\right)^i & = {}
    \frac{\left( \frac{q-1}{q} - \left(\frac{q-1}{q}\right)^{n}\right)}{1-\frac{q-1}{q}}\\
    & = {} (q-1) - q \left(\frac{q-1}{q}\right)^{n}.
\end{align}
Putting this in to Eq (\ref{appendixa2}), recognizing that $(q-1)^n=(-1)^n (1-q)^n$, and distributing the leading factor of $(-1)^{n+1}$, we get
\begin{equation}
    D_n =  \left(\frac{q}{1-q}\right)^{n/2}  \left[ \left(-1\right)^{n}(1-q) + q^{1-n} (1-q)^n \right],
\end{equation}
and this can be simplified to Eq (\ref{Dnrhoeqn}).

The starting point we used above was Eq (\ref{aminus}):
\begin{equation*}
    \frac{1-a_{-}/\mu}{\delta} = \left( \frac{q}{1-q} \right)^{1/2}.
\end{equation*}
If instead one focuses on $a_+$, the equivalent result is
\begin{equation}
    \frac{1-a_{+}/\mu}{\delta} = -\left( \frac{1-q}{q} \right)^{1/2}.
\end{equation}
Given that Eq (\ref{DNlimitsrepriseZ}) is unchanged when replacing $z \rightarrow -(1/z)$, the derivation holds whether using $a_+$ or $a_-$.  Thus, the `$a$' in Eq (\ref{DNlimitsreprise}) is valid for either meaning of $a$, and we have shown that Eqs (\ref{Dnrhoeqn}, \ref{DNlimits}) are equivalent.

\end{appendix}

\section*{Table of Symbols}
\begin{tabular}{ |c|c| } 
 \hline
 Symbol & Description \\
 \hline
 $\mu$ & mean \\ 
 $\sigma$ & standard deviation \\
 $\delta$ & coefficient of variation (CoV) = $\sigma/\mu$\\
 $m_n$ & $n$th central moment \\ 
 $D_n$ & $n$th standardized moment\\
 $M_n$ & convenient function of $D_n$ \\
 $x_{\rm extr}$ & the extreme values of $x$, both the minimum and maximum. \\
 $q$ & population fraction of large size in a bidisperse sample \\
$p$ & population fraction of small size in a bidisperse sample \\
$a_+$ & value of large size in a bidisperse sample \\
$a_-$ & value of small size in a bidisperse sample \\
$\eta$ & size ratio of bidisperse sizes \\
$\Delta'_+$ & difference of large size from mean \\
$\Delta'_-$ & difference of small size from mean \\
$\Delta_+$ & relative difference of large size from mean \\
$\Delta_-$ & relative difference of small size from mean \\
$P_r,P_s$ & two unique distributions \\
$P_t$ & combination of $P_r$ and $P_s$ in some proportionality \\
\hline
\end{tabular}

\section*{Acknowledgments}

We thank J. Crocker, Z. Germain, and P. Kragel for helpful discussions. 


\end{document}